# Correlating Nanocrystalline Structure with Electronic Properties in 2D Platinum Diselenide

*Sebastian Lukas, Oliver Hartwig, Maximilian Prechtl, Giovanna Capraro, Jens Bolten, Alexander Meledin, Joachim Mayer, Daniel Neumaier, Satender Kataria, Georg S. Duesberg, Max C. Lemme\**

Sebastian Lukas, Giovanna Capraro, Dr. Satender Kataria, Univ.-Prof. Dr.-Ing. Max C. Lemme
Chair of Electronic Devices, RWTH Aachen University, Otto-Blumenthal-Str. 2, 52074 Aachen, Germany
Email: max.lemme@eld.rwth-aachen.de

Oliver Hartwig, Maximilian Prechtl, Univ.-Prof. Dr. rer. nat. Georg S. Duesberg
Insitute of Physics, Faculty of Electrical Engineering and Information Technology (EIT 2), Universität der Bundeswehr München, Werner-Heisenberg-Weg 39, 85577 Neubiberg, Germany

Giovanna Capraro, Dr.-Ing. Jens Bolten, Univ.-Prof. Dr. rer. nat. Daniel Neumaier, Univ.-Prof. Dr.-Ing. Max C. Lemme
AMO GmbH, Advanced Microelectronic Center Aachen, Otto-Blumenthal-Str. 25, 52074 Aachen, Germany

Dr. Phys. Alexander Meledin, Univ.-Prof. Dr. rer. nat. Joachim Mayer
Central Facility for Electron Microscopy, RWTH Aachen University, Ahornstr. 55, 52074 Aachen, Germany

Dr. Phys. Alexander Meledin, Univ.-Prof. Dr. rer. nat. Joachim Mayer
Ernst Ruska-Centre for Microscopy and Spectroscopy with Electrons (ER-C), Forschungszentrum Jülich GmbH, 52425 Jülich, Germany

Univ.-Prof. Dr. rer. nat. Daniel Neumaier
Chair of Smart Sensor Systems, Bergische Universität Wuppertal, Lise-Meitner-Str. 13, 42119 Wuppertal, Germany

Platinum diselenide ($PtSe_2$) is a two-dimensional (2D) material with outstanding electronic and piezoresistive properties. The material can be grown at low temperatures in a scalable manner which makes it extremely appealing for many potential electronics, photonics, and sensing applications. Here, we investigate the nanocrystalline structure of different $PtSe_2$ thin films grown by thermally assisted conversion (TAC) and correlate them with their electronic and piezoresistive properties. We use scanning transmission electron microscopy for structural analysis, X-ray photoelectron spectroscopy (XPS) for chemical analysis, and Raman spectroscopy for phase identification. Electronic devices are fabricated using transferred PtSe$_2$ films for electrical characterization and piezoresistive gauge factor measurements. The variations of crystallite size and their orientations are found to have a strong correlation with the electronic and piezoresistive properties of the films, especially the sheet resistivity and the effective charge carrier mobility. Our findings may pave the way for tuning and optimizing the properties of TAC-grown PtSe$_2$ towards numerous applications.

## 1. Introduction

Layered platinum diselenide (PtSe$_2$) is a transition metal dichalcogenide (TMDC) with promising properties for novel electronic and sensing devices.[1–6] It can be sub-classified among the two-dimensional (2D) materials as a noble metal dichalcogenide (NMDC). It is a semiconductor with a bandgap of up to 1.6 eV in monolayer form and becomes a semimetal as the number of layers increases.[7–10] This metal-semiconductor transition may be utilized in lateral heterostructures with low-ohmic contacts of the semimetal phase to a channel of semiconducting PtSe$_2$ for electronics applications.[11–13] The predicted phonon-limited room-temperature charge carrier mobility of PtSe$_2$ is higher than 1000 cm$^2$ V$^{-1}$ s$^{-1}$ [14] with the highest experimentally measured value of 625 cm$^2$ V$^{-1}$ s$^{-1}$.[15] Furthermore, PtSe$_2$ films have been grown as both p-type and n-type semiconductors by varying the growth parameters,[16] which could pave the way for PtSe$_2$-based CMOS circuits. The semi-metallic nature of few-layer PtSe$_2$ suggests applications in infrared photodetection.[1,3,17,18] Finally, PtSe$_2$ has been demonstrated as material for highly sensitive membrane-based pressure sensors due to its high negative piezoresistive gauge factor (GF) of up to -84,[2] which is attributed to a change of the bandgap under mechanical strain.[19] The piezoresistive effect may be exploited in membrane-based nanoelectromechanical systems, with high potential for pressure, strain, acceleration or other sensors.[20]

In addition to the high application potential, PtSe$_2$ exhibits long-term stability in air[21] and it can be grown at complementary metal oxide semiconductor (CMOS)-back end of line (BEOL)-compatible temperatures (≤ 450 °C),[1,22] through thermally assisted conversion (TAC) of thin platinum films. Furthermore, TAC allows wafer-scale growth on various substrates, such as silicon dioxide (SiO$_2$),[15,22–25] aluminum oxide (Al$_2$O$_3$),[11] and other rigid substrates,[26] but also on flexible substrates.[15,27]

Even though the application potential of PtSe$_2$ has been clearly demonstrated, literature data of electronic properties shows high variability. Most studies report mobilities lower than 50 cm$^2$ V$^{-1}$ s$^{-1}$ [11,16,25,28–31] with some lower than 1 cm$^2$ V$^{-1}$ s$^{-1}$ [12,23,32] for TAC-grown PtSe$_2$. Also, semiconducting behavior has been reported for much thicker films than the theoretically expected mono- and bilayers.[25] The p- and n-doping of PtSe$_2$ films is reported to depend on the selenization process.[16] Further we have also found significant variations in the piezoresistive gauge factor in our earlier report.[27] Thus, PtSe$_2$ and its electronic properties appear to be highly tunable.

TAC-grown 2D materials are mostly polycrystalline materials with thicknesses of only few atomic layers and domains in the nanometer scale.[1,8,27,33,34] Similar to other 2D materials such as graphene[35,36] or MoS$_2$,[37,38] the resistivity and the effective mobility are influenced by the sizes of individual crystallites and their respective orientation. Horizontally aligned layers for very thin TAC-grown films have repeatedly been reported, while thicker films have shown to possess vertically aligned layers on the substrates.[39,40] Thus, correlating the individually reported basic material properties with electronic performance in a systematic way remains challenging.

Here, we present a detailed study of PtSe$_2$ films grown by TAC at 450 °C and find a strong correlation between their microstructures and measured electronic and piezoresistive properties. Cross-sectional annular bright field and annular dark field scanning transmission electron microscopy (ABF/ADF STEM) is utilized for a thorough analysis of their microstructures.

Detailed Raman analysis reveals that the peak width of the $E_g$ mode is a suitable indicator for the film quality. Furthermore, X-ray photoelectron spectroscopy (XPS) is used for chemical analysis. Finally, we measure and analyze the electronic properties and the piezoresistive gauge factor (GF) on devices with transferred PtSe$_2$ films. We then discuss the structural composition and variations in the nanocrystalline films and present a model correlating the polycrystalline electrical resistivity and its change with temperature[41,42].

## 2. Results and discussion

PtSe$_2$ films of different thicknesses have been grown on silicon (Si) / SiO$_2$ and quartz substrates by TAC in different deposition chambers and in four different batches (see **Table 1** for an overview). For electronic device fabrication, the films of sample batches 1, 2, and 3 have been wet-transferred from their growth substrates onto highly p-doped (p+) Si / 90 nm SiO$_2$ substrates. The films of sample batch 4 had been directly grown on p+ Si / 90 nm SiO$_2$ substrates and electronic device fabrication and characterization was therefore done on the growth substrate without transfer. To verify that the transfer did not significantly modify the films, some characterization was also done on transferred films of batch 4. Films of all batches have additionally been transferred onto flexible polyimide foil (Kapton) substrates for strain gauge fabrication. Details on the material synthesis and transfer processes can be found in the experimental section.

### 2.1 Material analysis

All samples were characterized using Raman spectroscopy, which provides information on their phase formation and crystallinity. The recorded spectra show the characteristic peaks of layered PtSe$_2$ at approx. 175 cm$^{-1}$ ($E_g$ peak) and 205 cm$^{-1}$ ($A_{1g}$ peak)[8] in all examined samples (**Figure 1**a). The $E_g$ peak positions shift to lower wave numbers for increasing film thicknesses (Figure 1b, bottom), in agreement with previous findings.[3,8,9,13,33,43,44] The shifts of the $A_{1g}$ peak positions, however, do not display an obvious dependence on the film thickness (Figure 1b, top). Furthermore, the intensity ratio of the two peaks, $I(A_{1g})/I(E_g)$, generally increases with increasing film thicknesses (Figure 1c), which is in line with experimental studies[8,9,43,44] and theoretical calculations.[45] For ideal bulk $PtSe_2$, an intensity ratio approaching 1 is expected. Here, we measured ratios close to 0.5 up to a film thickness of 15 nm, which increased to close to 0.7 for a film thickness of 23 nm. The intensity ratios vary for samples of the same thickness. The measured intensity ratios are lower than expected for bulk $PtSe_2$, suggesting that the polycrystalline TAC-grown $PtSe_2$ films consist of stacked crystallites of only very few layers each, producing a superposition of few-layer Raman signatures, with some variation between samples.

It was previously suggested that the full width at half maximum (FWHM) of the $E_g$ peak in $PtSe_2$ is an indicator of the film quality,[33,44] similar to the FWHM of the 2D peak in graphene correlating to its charge carrier mobility.[46,47] An $E_g$ FWHM of $\leq 7$ cm$^{-1}$ has been proposed as an indicator for high quality layered films.[44]

Histograms of the FWHM of the $E_g$ peak were extracted from Raman area scans (Figure 1d). The FWHM varied between approximately 4.5 cm$^{-1}$ and 8.5 cm$^{-1}$ across the different samples, independent of the film thickness.

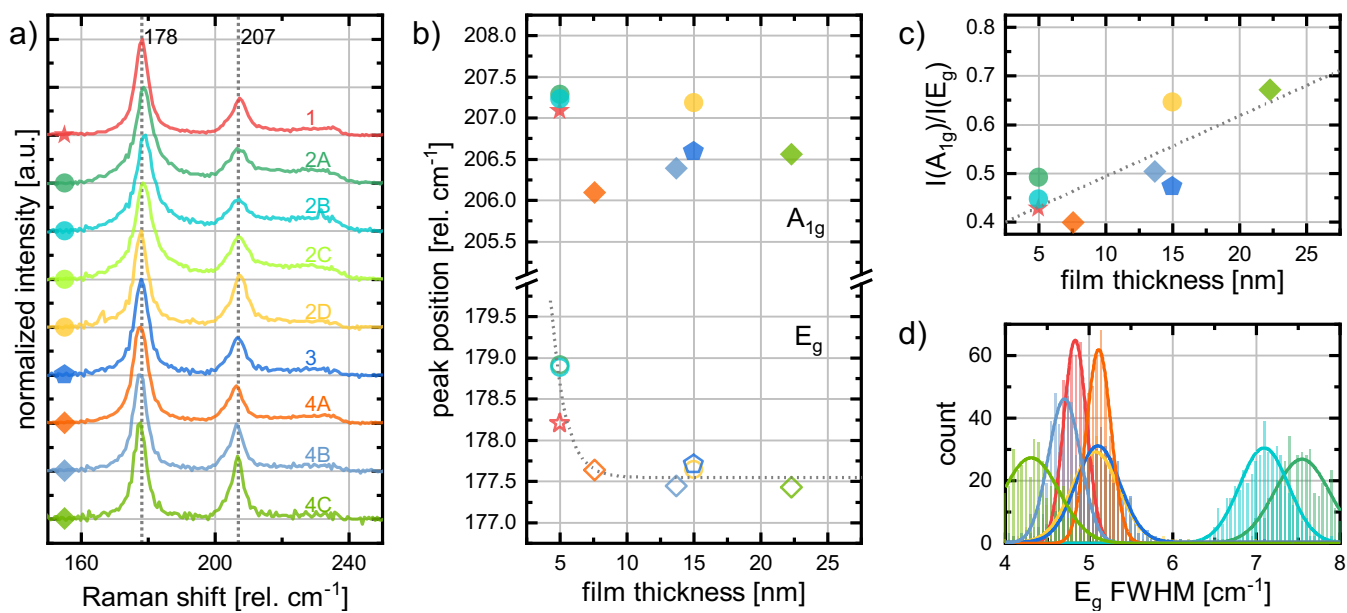

**Figure 1.** Raman analysis of the various PtSe$_2$ films. (a) Raman spectra of nine of the examined PtSe$_2$ samples. The spectra were recorded with an integration time of 8 s and averaged over 10 accumulations. The data was then normalized to the maximum intensity of the E$_g$ peak. For better comparison, two dotted reference lines at 178 cm$^{-1}$ and 207 cm$^{-1}$ are shown. (b) Shift of the E$_g$ peak (hollow symbols) and A$_{1g}$ peak (solid symbols) position depending on the PtSe$_2$ film thickness. (c) Intensity ratio of the two characteristic Raman peaks of PtSe$_2$ over the approximate film thickness. (d) Histograms of the E$_g$ peak FWHM from Raman area scans. In all panels, the dotted lines are just a guide to the eye. Sample numbers are indicated in panel (a) and in Table 1.

A comparison of Raman spectra of as-grown and transferred samples of batch 4 shows no major differences in peak position and peak width (SI **Figure S1**). This indicates that the transfer process had no impact on the film quality.

The samples have been further analyzed with x-ray photoelectron spectroscopy (XPS) for revealing their chemical composition. The results show no significant differences in peak features (SI **Figure S2**), indicating similar chemical compositions of all investigated films. Indeed, the extracted atomic ratios of Se and Pt atoms were approximately 1.7 for samples 1, 2A, 2D, and 3 and approximately 1.8 for samples 4A, 4B, and 4C with deviations within the measurement tolerances. Significant differences in the chemical composition of the films like oxygen content or unselenized Pt can, therefore, be ruled out.

The PtSe$_2$ films of samples 1, 2A, 2D, 3, 4A, and 4C were analyzed using annular bright field and annular dark field scanning transmission electron microscopy (ABF/ADF STEM). Cross-sectional images of the films reveal a polycrystalline structure with varying crystallite sizes and arrangements (see **Figure 2** and SI **Figure S3**). The STEM images of films with similar thickness, such as samples 1 and 2A with 5 nm thickness, reveal significant differences in their nanocrystalline structure. For example, the images of sample 1 display a smooth arrangement of the PtSe$_2$ crystallites, highly oriented and parallel to the substrate (Figure 2a and SI Figure S3a). The crystallites lateral dimensions are found to be approximately 10 nm in size. The images of sample 2A, in contrast, reveal a film with much rougher surface and randomly oriented crystallites with lateral dimensions of approximately 5 nm (Figure 2b and SI Figure S3b). The images of sample 4A also show a rather smooth surface, similar to sample 1, albeit slightly smaller crystal sizes (Figure 2e and SI Figure S3e).

The STEM images of the thicker films of samples 2D, 3, and 4C, show the atomic nanocrystalline layers only close to the SiO$_2$ substrate (Figure 2c, d, f and SI Figure S3c, d, f). Crystallites in the upper parts of the films may be tilted out of the plane of the substrate and their atomic layer structure therefore becomes invisible for STEM. In the thickest of all examined films in sample 4C, several almost vertically grown crystallites are visible. Vertically aligned layers of PtSe$_2$ and other 2D dichalcogenides have been reported previously for thicker films.[15,48,49] The STEM investigations confirm the structural variations as indicated indirectly by the Raman spectroscopic studies discussed earlier.

When comparing the lateral dimensions of the crystallites across the different samples, sample 4C exhibits the largest crystallites with more than 10 nm lateral dimension, coinciding with the lowest Raman E$_g$ peak FWHM measured for this sample. Samples 1, 4A, 2D, and 3 follow with similar, slightly larger E$_g$ FWHM and similar, slightly smaller estimated crystallite sizes, and finally sample 2A shows the highest E$_g$ FWHM while clearly possessing the smallest crystallites.

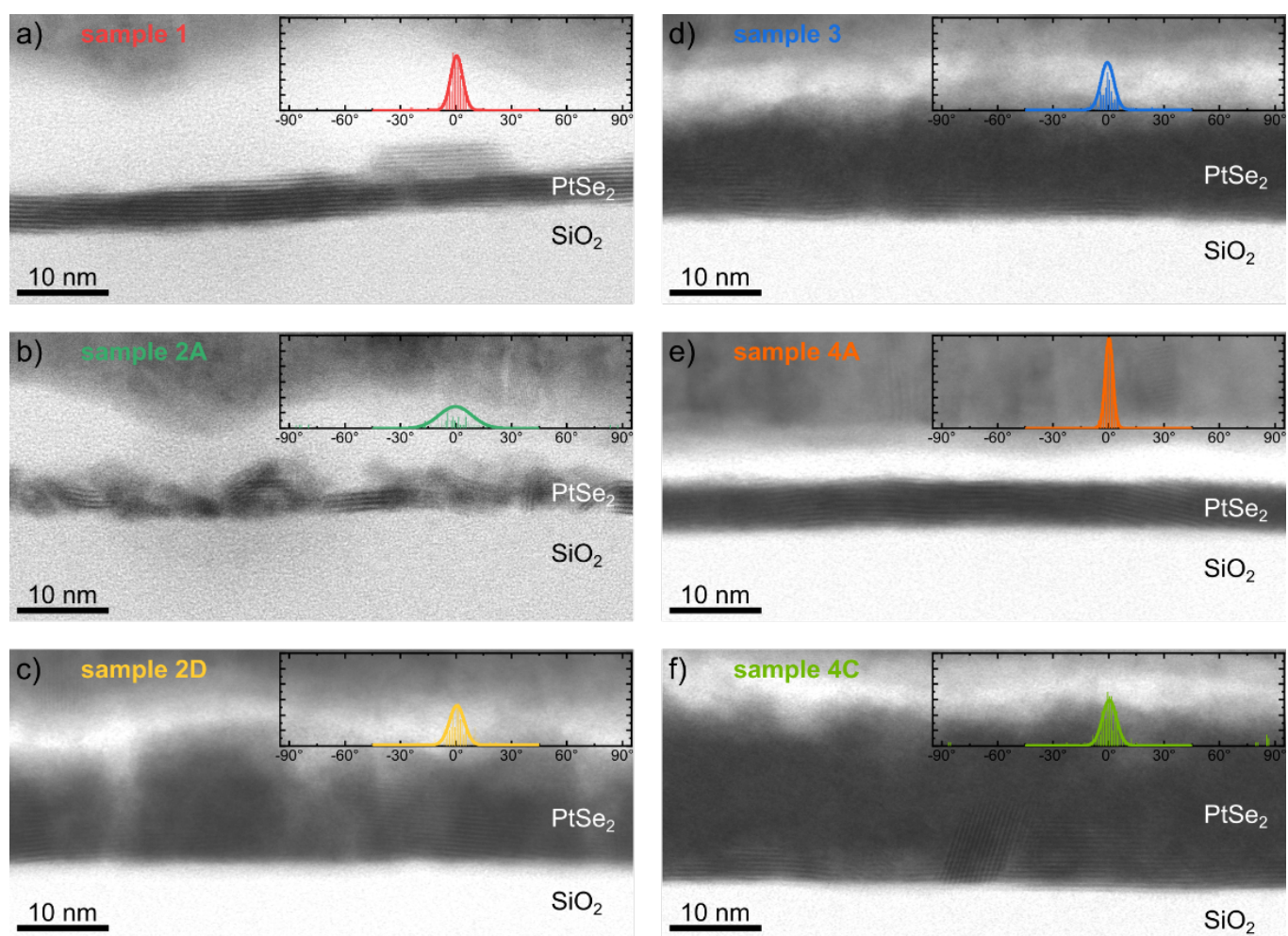

**Figure 2**. Annular bright field scanning transmission electron microscopy (ABF STEM) images of the PtSe$_2$ films of (a) sample 1, (b) sample 2A, (c) sample 2D, (d) sample 3, (e) sample 4A, and (f) sample 4C. Additional images at slightly higher magnification can be found in SI **Figure S3** and **S4**. The insets show histograms with Gaussian fits of the distribution of the crystallite tilting angles (on x-axis) extracted from annular dark field (ADF) STEM image analysis (see also SI **Figure S6**).

While the previous analysis was only based on the observation of individual ABF STEM images by naked eye, several images of all six samples were quantitatively analyzed using a pattern recognition feature of a software tool dedicated to automated data extraction from images (GenISys ProSEM[50]). For this purpose, the image contour lines were enhanced by suitable image filters (using ImageJ software) to facilitate the pattern recognition. Several 5 nm square sections were selected within each PtSe$_2$ image, which were then analyzed by an algorithm that yields the tilting angle of the crystallites with respect to the SiO$_2$ surface within each image section. The analysis of the extracted angles in 585 sections across 57 ADF STEM images resulted in the statistical distribution of tilting angles of the crystallites (see insets in Figure 2a-f). A clear difference is seen between samples 1 and 2A, which also showed the largest difference in electronic properties (see below). The tilting angles for sample 2A are much more spread than for sample 1 (standard deviation: 30.59 ° vs. 6.51 °). Furthermore, some almost vertically oriented crystallites can be identified in the angle distributions for samples 2A and 4C (not included in the fit to determine the standard deviation), as previously observed in the STEM images. Automating the analysis of the STEM images has several uncertainties. Crystallites tilted out of the plane of view are not visible in STEM images and are therefore not included in the analysis. In addition, the number of analyzed image sections varies across the six examined samples, because images of well-aligned crystallites are more easily registered and subsequently analyzed by pattern recognition. Nevertheless, the statistical analysis underlines the observation of increased nanocrystalline disorder in several of the $PtSe_2$ films.

## 2.2 Device fabrication and characterization

After structural characterization, electronic devices based on transferred $PtSe_2$ films were fabricated and characterized to explore their electronic and piezoresistive properties.

Transfer length method (TLM) and six-port Hall bar structures have been defined with optical contact lithography and the $PtSe_2$ films have been patterned using $CF_4/O_2$-based reactive ion etching (RIE). Nickel/aluminum (Ni/Al) edge contacts have been realized through a self-aligned method.[51] The highly doped Si substrate acts as a global back gate with the 90 nm $SiO_2$ layer as gate dielectric. A cross-sectional schematic of such devices is shown in **Figure 3**a. SEM images of the six-port and TLM devices are shown in Figure 3c and Figure 3d, respectively. Raman scans across the patterned device channels reveal the successful patterning of $PtSe_2$ device channels without noticeable damage (see Figure 3b).

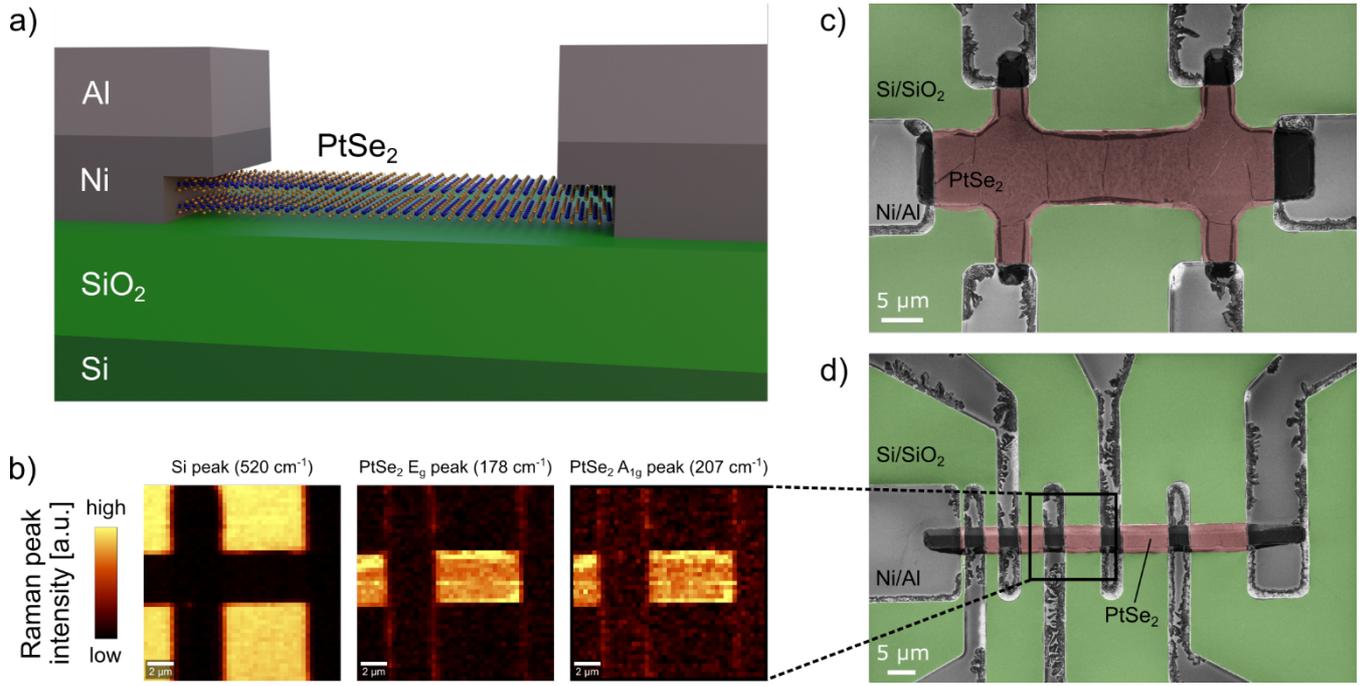

**Figure 3**. (a) Schematic cross section of the PtSe$_2$ devices. (b) Raman maps of the intensities of the Si peak (520 cm$^{-1}$), and the E$_g$ and A$_{1g}$ peaks of PtSe$_2$ after etching the PtSe$_2$ for channel patterning of a TLM device. The bright, yellow pixels represent high intensity of the respective peak, while the dark pixels represent low intensity. (c) False-color SEM images of a six-port device and (d) a TLM device. The PtSe$_2$ channel is shown in red and the SiO$_2$ substrate is shown in green.

TLM measurements have been performed to extract the contact resistance ($R_c$) and the sheet resistance ($R_\square$) of the various samples without gate bias (SI **Figure S7**). The sheet resistance has additionally been measured in four-point measurements using the six-port Hall bar devices. The $R_\square$ data show a dramatic spread over several orders of magnitude between the different samples. For instance, samples with similar film thicknesses of 5 nm exhibit $R_\square$ variations of approximately four orders of magnitude, i.e. between 6 kΩ/□ and more than 100 MΩ/□. $R_c$ was difficult to precisely extract for the high-ohmic samples 2A, 2B, and 4A due to large scattering of the data points. For all samples with $R_\square$ < 50 kΩ/□, $R_c$ was between 0.7 kΩμm and 2.1 kΩμm.

The six-port Hall bar structures have been used to conduct back-gated field-effect measurements in a four-point set-up (**Figure 4**c). The four-point configuration allows the extraction of the charge carrier mobility ($\mu$) without the parasitic influence of $R_c$,[52–56] which varied greatly in the TLM measurements. It is based on the two-point method, where the effective field-effect charge carrier mobility ($\mu_{2P}$) can be calculated according to **Equation (1)**[57,58],

$$\mu_{2P} = \frac{\partial I_D}{\partial V_{BG}} \cdot \frac{1}{V_{DS}} \cdot \frac{L}{W} \cdot \frac{d_{ox}}{\varepsilon_{r,ox}\varepsilon_0} \qquad (1)$$

Here, $I_D$ is the bias-driven current through the channel, $V_{BG}$ is the back-gate voltage, $V_{DS}$ is the voltage applied to the channel, $L$ is the channel length between the two contacts, $W \approx 10$ μm is the channel width, $d_{ox}$ = 90 nm is the gate oxide thickness, $\varepsilon_{r,ox}$ = 3.9 is the relative permittivity of the gate oxide (SiO$_2$), and $\varepsilon_0$ is the vacuum permittivity. In the four-point configuration, the two inner contacts are used to measure the differential voltage between the two neighboring inner contacts ($V_{diff}$), which replaces $V_{DS}$ in Equation (1). Since $V_{diff}$ may change with $V_{BG}$, it must be placed in the derivative. The effective field-effect charge carrier mobility ($\mu_F$) in the four-point set-up can then be extracted according to **Equation (2)**,

$$\mu_F = \frac{\partial(I_D/V_{diff})}{\partial V_{BG}} \cdot \frac{L_{inner}}{W} \cdot \frac{d_{ox}}{\varepsilon_{r,ox}\varepsilon_0} \qquad (2)$$

This extraction method leads to a macroscopic value of the effective mobility of the polycrystalline material, whereas the intrinsic mobility of individual single-crystal PtSe$_2$ will certainly be higher, as reported from experimental studies on exfoliated PtSe$_2$.[13,21] All devices showed p-type behavior as their resistance decreased with negative gate bias (Figure 4a). Large variations in $\mu_F$ were measured, ranging from below 0.005 cm$^2$ V$^{-1}$ s$^{-1}$ (samples 2A and 2B) up to more than 13 cm$^2$ V$^{-1}$ s$^{-1}$ (sample 1) on samples of similar PtSe$_2$ film thickness. Note that for samples 2A and 2B, the gate and the drain current were of similar magnitude (< 0.5 nA), as shown in SI **Figure S8**. Therefore, the corresponding mobility measurements must be treated with caution. In general, the lowest mobilities were measured in films of poor crystalline quality, according to the TEM and Raman analysis. The $R_\square$ values are plotted as a function of the extracted $\mu_F$ in Figure 4b, which illustrates an obvious correlation between the two macroscopic values.

Note that for batch 4, $R_\square$, $R_c$, and $\mu_F$ have been extracted from devices of both as-grown and transferred films with no significant differences observed. As already seen from the Raman spectra of batch 4 samples, the transfer method does therefore not significantly modify the PtSe$_2$ film quality.

The extracted electronic properties like sheet resistance and carrier mobility correlate well with the E$_g$ peak FWHM as retrieved from Raman analysis, with an exponential and inverse exponential dependence, respectively (Figure 4d and e). Film thickness alone is thus clearly not a suitable measure to predict the electronic properties of TAC-grown PtSe$_2$. Raman spectroscopy can provide additional insights into expected electronic properties of PtSe$_2$. However, judging from our experiments, high quality TAC-grown PtSe$_2$ films should exhibit an E$_g$ FWHM of < 5 cm$^{-1}$.

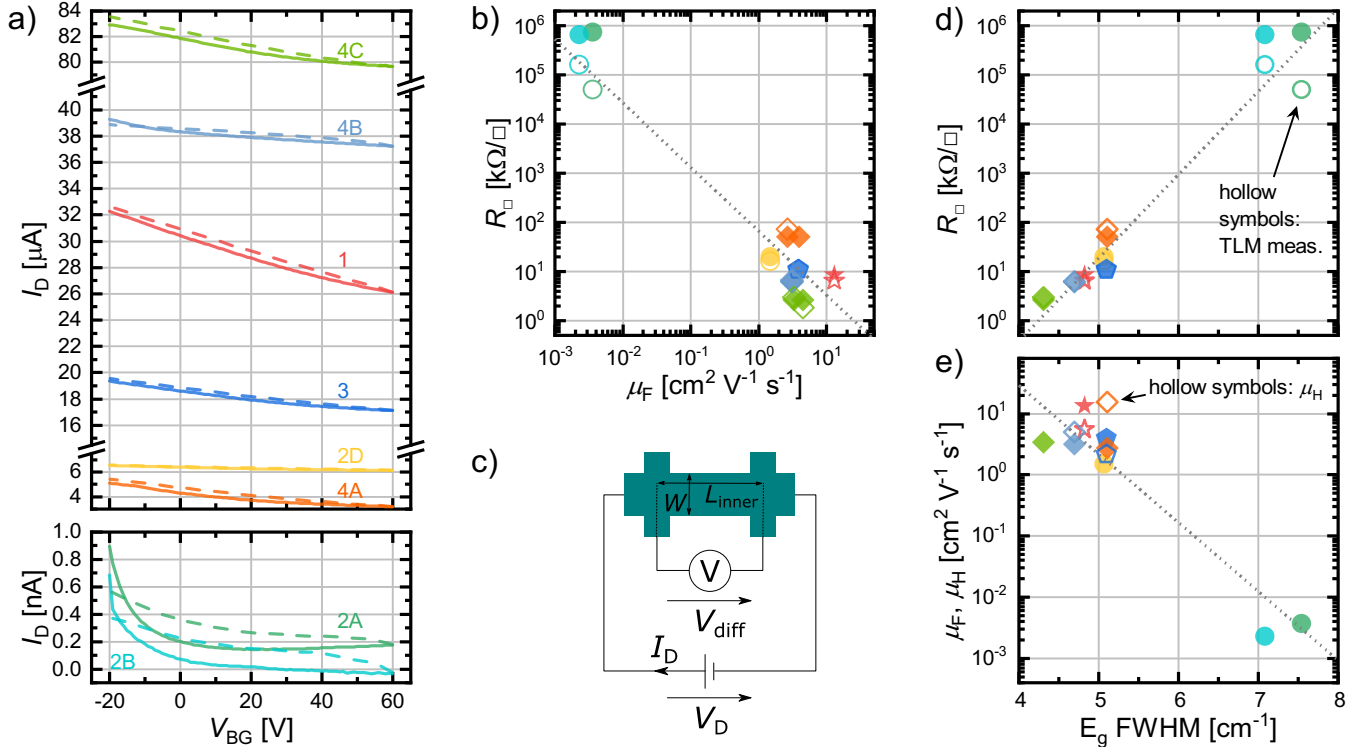

**Figure 4**. (a) Field-effect measurements of the various PtSe$_2$ devices. The back-gate voltage $V_{BG}$ was swept from -20 V to 60 V (solid line) and back (dashed line). Sample numbers are indicated. (b) Correlation between the field-effect mobility $\mu_F$ and the sheet resistance $R_\square$. PtSe$_2$ samples with a high $R_\square$ exhibit a low $\mu_F$ and vice versa. The filled and hollow symbols correspond to four-point and TLM measurements, respectively. (c) Schematic of the four-point set-up used for field-effect mobility and sheet resistance extraction. (d) Exponential dependence of the sheet resistance $R_\square$ on the Raman E$_g$ peak FWHM. The filled and hollow symbols correspond to four-point and TLM measurements, respectively. (e) Dependence of the field-effect mobility $\mu_F$ (solid symbols) and the Hall mobility $\mu_H$ (hollow symbols) on the Raman E$_g$ peak FWHM. In all panels, the dotted lines are just a guide to the eye. Sample numbers are color-coded in panel (a) and indicated in **Table 1**.

AC-modulated Hall effect measurements were performed on four samples with sufficiently high $\mu_F$ (sample numbers 1, 3, 4A and 4B), since the generally low effective mobilities did not result in Hall signals in a standard DC Hall set-up. Hall voltages on the order of a few 100 µV have been extracted from the AC Hall signal (see SI **Figure S9**b), which corresponds to effective Hall mobilities between 2.1 cm$^2$ V$^{-1}$ s$^{-1}$ and 15.3 cm$^2$ V$^{-1}$ s$^{-1}$ (see Table 1 for details). The extracted sheet charge carrier densities ($n_\square$) were between 7.3 × 10$^{12}$ cm$^{-2}$ and 2.9 × 10$^{14}$ cm$^{-2}$ in line with previous Hall data in PtSe$_2$ devices.[22] The polarity of the measured Hall voltages confirmed that the PtSe$_2$ samples were p-type. However, we cannot rule out the presence of both holes and electrons in the samples, which is why the reported $\mu_H$ and $n_\square$ should be treated with caution.

The temperature-dependence of the resistance can give insights into the details of the charge transport inside the material. $R_\square$ of several PtSe$_2$ devices on SiO$_2$ substrates was therefore measured at predefined temperatures up to 100 °C in ambient conditions in the four-point set-up. All devices displayed a decrease of $R_\square$ with increasing temperature, typical for temperature-dependent carrier transport in semiconducting crystals[24] and similar to previously observed behavior in PtSe$_2$ at a lower temperatures.[26] This is an indication that the nanocrystalline PtSe$_2$ films are of semiconducting nature, at least partially. Note that the resistance values went back to their original level after cool-down. The data points were fitted to a linear curve according to **Equation (3)**,[59]

$$R_\square = R_{\square,0} \left(1 + \alpha_{\text{lin}} (T - T_0)\right) \tag{3}$$

where $R_{\square,0}$ is the sheet resistance at the reference temperature $T_0 = 25$ °C. The data is shown in SI **Figure S10**a. The temperature coefficient of resistance, $\alpha_{\text{lin}}$, was determined to range from zero (i.e. no variation with temperature) to approximately -0.012 K$^{-1}$ for measurements without applying a back-gate voltage. By absolute value, devices with the lowest $R_\square$ had the lowest $\alpha_{\text{lin}}$, and the absolute value of $\alpha_{\text{lin}}$ increased towards higher $R_\square$. This trend is observed over a $R_\square$ range of almost six orders of magnitude (see SI Figure S10b). An Arrhenius plot in **Figure 5**a was fitted with **Equation (4)**,

$$\ln\left(\frac{R_\square}{1\text{ k}\Omega/\square}\right) = m\,\frac{1}{T} + y_0 \tag{4}$$

Here, $m$ is the slope and $y_0$ is the y-axis intercept of the fit. From the slope, an activation energy ($E_A$) can be calculated according to $m = E_A/k_B$ where $k_B$ is the Boltzmann constant. In this way, $R_\square$ follows as **Equation (5)**,

$$R_\square = R_{\square,\infty}\, e^{\frac{E_A}{k_B T}} \tag{5}$$

with the fitting parameter $R_{\square,\infty} = 1$ k$\Omega/\square \times e^{y_0}$. We interpret $E_A$ as a measurement of the intergrain barrier height. A clear correlation with the extracted $R_\square$ is evident (Figure 5b), very similar to the relationship of $\alpha_{lin}$ and $R_\square$. The extracted $E_A$ ranges from zero to 32 meV for all low-resistance samples ($R_\square <$ 100 k$\Omega/\square$) and increases up to 160 meV for the extremely high-resistance samples with almost gigaohm-range $R_\square$. These higher values are in agreement with previous work on transferred TAC-grown PtSe$_2$, where an activation energy of approximately 200 meV was extracted at zero gate voltage[22] and with work on pristine TAC grown PtSe$_2$, where an activation energy of 134 meV was reported.[24] In the latter, a decrease of $E_A$ is attributed to an Ar plasma treatment and a resulting reduction of the Se content in their samples. This is in contrast to our samples, where the ratio of Pt and Se was determined to be approximately the same in XPS measurements for samples of very different $E_A$. Furthermore, we observe a decrease of $E_A$ in the ON state, i.e. at negative back-gate voltages (SI **Figure S11**), for both as-grown and for transferred films. Previously, this effect has been observed only after transfer of PtSe$_2$, but not for as-grown films.[22] The effect is small for samples 4B and 4C, but more pronounced for sample 4A, where $E_A$ changes from 25 meV in the ON-state to 45 meV in the OFF-state. However, our non-transferred samples generally exhibit a lower $E_A$ than those in [22].

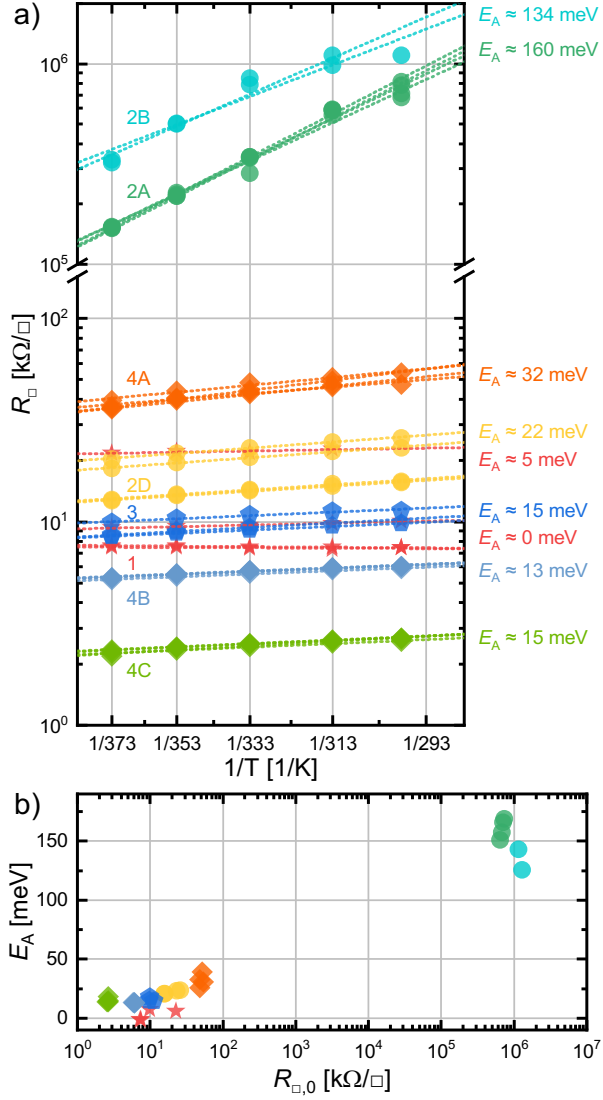

**Figure 5**. Temperature dependence of the sheet resistance $R_\square$. (a) Arrhenius plot of the temperature dependence of $R_\square$ of 29 measured devices. The measurements were done without applied back-gate voltage (floating gate). An exponential dependence of $R_\square$ on the inverse temperature is seen, which is a sign of an energy barrier that the charge carriers need to overcome. We suggest that this barrier may be caused by the grain boundaries in the polycrystalline film. Fits are shown as dotted lines and average extracted activation energies $E_A$ are shown for each sample. (b) Dependence of the extracted activation energy $E_A$ on the initial sheet resistance at $T = 25$ °C. Sample numbers are indicated in panel (a) and in Table 1.

The piezoresistive gauge factor (GF) of PtSe$_2$ is an important parameter for membrane-based sensor applications. The GF was determined using unpatterned PtSe$_2$ films transferred to flexible substrates glued to a steel beam with an attached mass. The set-up is described in detail in the methods section and in [2]. The direction of current flow through the PtSe$_2$ films was parallel to the direction of mechanical strain due to the bending of the steel beam. A constant DC voltage was applied while the total resistance of the films was monitored during loading and unloading of the beam with a mass of 2 kg. Resistance changes upon mechanical strain were evident for all PtSe$_2$ films, although their magnitude and even their sign varied. The GF can generally be calculated with Equation (6),[60]

$$\text{GF} = \frac{\Delta R}{R_0 \varepsilon} = 1 + 2\nu + \frac{\Delta \rho}{\rho_0 \varepsilon} \tag{6}$$

where $\Delta R$ is the difference between the initial resistance $R_0$ without strain and the resistance under the strain $\varepsilon$ at the position of the sample. Furthermore, $\nu$ is Poisson's ratio, $\Delta \rho$ is the change in resistivity from the initial resistivity $\rho_0$ under the strain $\varepsilon$. Therefore, $1 + 2\nu$ is the geometric contribution and $\Delta \rho / \rho_0 \varepsilon$ is the piezoresistive contribution to GF. Tensile (positive) or compressive (negative) strain can be applied by mounting the sample on the top or bottom of the beam, respectively. For an applied mass of 2 kg, the strain is $\varepsilon \approx \pm 4.4 \times 10^{-4} = \pm 0.044\,\%$, as calculated in [2]. In the following, the average GF from measurements with tensile and compressive strain are reported. As with the resistance and mobility values, the spread in GF is large, ranging from -64.9 to +13.5 across seven measured samples. While single-digit positive GF are typical for metals such as aluminum, copper, gold, iron, platinum, or silver,[61] and originate mainly from the geometrical contribution of the equation, the piezoresistive contribution can dominate the equation in semiconductors and lead to a negative GF. In the latter case, changes in the band structure cause changes in the mobility and the charge carrier density, for example in silicon or germanium, where the GF can vary from less than -150 to more than +150, depending on the crystalline orientation and doping.[61] Due to the high negative GFs observed in some of the present polycrystalline PtSe$_2$ films, they are likely to include semiconducting crystallites, in line with the evidence provided through the temperature dependent transport measurements. In addition, the arrangement of the crystallites in the film is expected to have an influence on the GF, as the applied strain can change the grain overlap area through small shifts of the crystallite positions, similar to graphene flakes on graphene-ink coated substrates.[62]

In **Figure 6**b, the measured $R_\square$ is plotted versus the GF. No clear correlation can be observed, in particular for samples with low sheet resistance. The situation is different when plotting the carrier mobility versus the GF (Figure 6c). Here, higher negative GFs are generally observed for higher mobility samples, in particular when considering Hall mobility data. The field effect mobility data of samples 4A, 4B, and 4C does not fit this trend, as they show different, negative and positive GFs at similar $\mu_F$, although their growth process only differs in the thickness of the initial Pt layer. For samples 4A, 4B, and 4C, the GF increases with film thickness, a trend that is not generally confirmed by our data. In contrast, distinctly different electronic and piezoresistive behavior have been observed in samples of similar thickness as well as growth and processing conditions. Samples 1 and 2A in particular show variations by many orders of magnitude.[63]

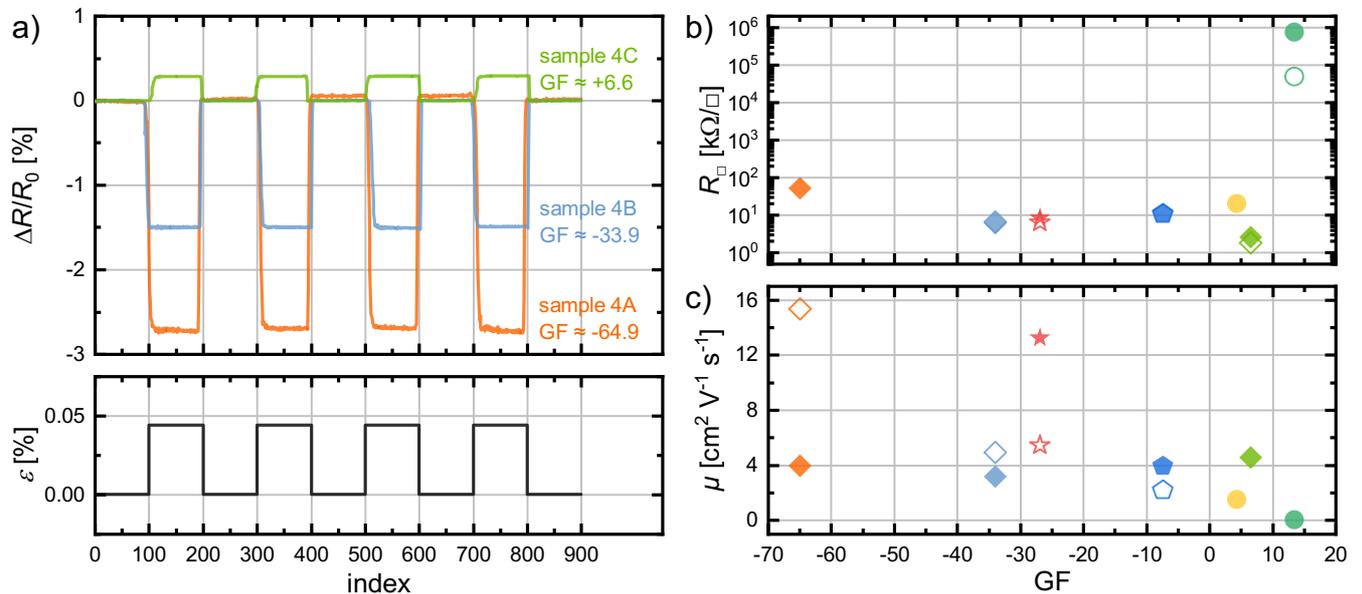

**Figure 6**. (a) Measurements of the change of device resistance $\Delta R/R_0$ under applied strain over time, shown for three samples of material batch 4. The bottom panel shows the applied tensile strain $\varepsilon$. One curve was recorded within approx. 2 minutes. (b) Correlation of the piezoresistive gauge factor (GF) and the sheet resistance $R_\square$. The filled and hollow symbols correspond to four-point and TLM measurements, respectively. (c) Correlation of GF and the extracted effective charge carrier mobility $\mu$. Here, the filled and hollow symbols correspond to $\mu_F$ and $\mu_H$, respectively. Sample numbers are indicated in Table 1.

The STEM images do not present a full explanation to the variation of the GF. It appears that the GF is influenced by both the total film thickness and the nanocrystalline structure. We observe high negative values of the GF for well-aligned, thin films of PtSe$_2$ (< 10 nm, samples 1 and 4A), and low positive values of the GF for generally thicker and less well aligned films (samples 4C and 2D). A thin but very disordered film (sample 2A), however, results in a positive GF as well.

The correlation between GFs and the electronic and material properties in TAC is a complex issue and cannot be explained by assuming ideal 2D layered PtSe$_2$. In addition, the GF is related to the change of the product of charge carrier density and mobility with strain in conventional, three-dimensional semiconductors,[61] data that could not be extracted from the experiments conducted in this work.

**2.3 Modelling**

The correlation between the electronic properties and the polycrystalline structure of a material can be explained through the size and number of the grain boundaries, acting as tunnel barriers for the charge carriers. Experience from research on other 2D materials such as graphene[35,36] or MoS$_2$[37,38] suggests that the resistivity and the effective mobility are influenced by the sizes of individual crystallites and their respective orientation. However, these materials are commonly grown by chemical vapor deposition yielding grains on micrometer scale or larger. Therefore, grain boundaries are likely to make up a significant part of the film and therefore can be expected to play a major role in determining the electronic properties, competing with the intrinsic electrical properties of the material.

Assuming similar chemical composition of the films and negligible differences during device fabrication, two properties influence $R_\square$. One factor is the average grain size, correlating with the number of grain boundaries in the film. A second factor is the barrier height presented by the grain boundaries. In our PtSe$_2$ films, the effects can be captured in a resistivity model for polycrystalline films,[41,42] which qualitatively captures the experimental observations of an increase in resistivity for a film with both decreasing crystallite size and increasing grain boundary density. The model is described in detail in the SI (**Figure S13**).

The change of resistivity upon heating can also be understood with the above-mentioned model. From SI Figure S13b, it is evident that even a minor change in the reflectivity coefficient (*r*) of the grain boundary barrier, possibly enforced through an increase in kinetic energy of the charge carriers, i.e. a higher temperature of the polycrystalline film, can greatly influence the polycrystalline resistivity. This underlines the significance of controlling the nanocrystalline structure, especially the amount and size of grain boundaries, in TAC-based growth processes.

**Table 1.** Overview of the experimental results from the samples studied within this work. $R_{\square,4P}$ and $R_{\square,TLM}$ are the sheet resistance values determined from four-point and TLM measurements, respectively. $R_c$ is the contact resistance determined from TLM measurements. $\mu_F$ and $\mu_H$ are the effective charge carrier mobility values determined from four-point and Hall measurements, respectively. $n_{\square}$ is the sheet carrier concentration extracted from Hall measurements. $E_A$ is the activation energy extracted from high temperature measurements of the device resistance without gate bias. GF is the piezoresistive gauge factor. $E_g$ FWHM is the width of the $E_g$ peak determined in Raman spectroscopy. - denotes that the data was not available, * denotes that the thickness was determined from AFM measurements, and · denotes that the thickness was estimated from the TEM images; other thicknesses are estimated from the initial Pt thicknesses.

| material batch | growth substrate | symbol | sample number | approx. thickness [nm] | $R_{\square,4P}$ [kΩ/□] | $R_{\square,TLM}$ [kΩ/□] | $R_c$ [kΩμm] | $\mu_F$ [cm² V⁻¹ s⁻¹] | $\mu_H$ [cm² V⁻¹ s⁻¹] | $n_{\square}$ [cm⁻²] | $E_A$ [meV] | GF | $E_g$ FWHM [cm⁻¹] |
|---|---|---|---|---|---|---|---|---|---|---|---|---|---|
| 1 | Si/SiO₂ | ★ | 1 | 5 · | 8.3 | 6.3 | 1.7 | 13.2 | 5.4 | 2.1 × 10¹⁴ | ≤ 5 | -26.8 | 4.83 |
| 2 | quartz | ● | 2A | 5 · | 748,000 | 50,000 | 32,000 | 0.0036 | - | - | 160 | 13.5 | 7.54 |
| 2 | quartz | ● | 2B | 5 | 667,000 | 160,000 | - | 0.0023 | - | - | 134 | - | 7.09 |
| 2 | quartz | ● | 2C | 7.5 | - | - | - | - | - | - | - | - | - |
| 2 | quartz | ● | 2D | 15 · | 20.1 | 16.0 | 1.3 | 1.5 | - | - | 22 | 4.4 | 5.07 |
| 3 | quartz | ⬟ | 3 | 15 · | 10.8 | 10.5 | 0.97 | 3.9 | 2.1 | 2.9 × 10¹⁴ | 15 | -7.3 | 5.11 |
| 4 | Si/SiO₂ | ◆ | 4A | 7.6 * | 50.7 | 71.1 | - | 2.7 | 15.3 | 7.3 × 10¹² | 32 | -64.9 | 5.12 |
| 4 | Si/SiO₂ | ◆ | 4B | 13.7 * | 6.0 | 6.1 | 2.1 | 3.1 | 4.9 | 2.3 × 10¹⁴ | 13 | -33.9 | 4.71 |
| 4 | Si/SiO₂ | ◆ | 4C | 22.3 * | 2.6 | 2.9 | 0.7 | 3.3 | - | - | 15 | 6.6 | 4.32 |

## 3. Conclusion

We have demonstrated that the electronic properties of TAC-grown PtSe$_2$ films greatly depend on the nanocrystalline structure of the materials. Correlations between structure and electronic properties are discussed based on detailed Raman and TEM studies and electrical measurements. Cross-sectional TEM images reveal that TAC-grown PtSe$_2$ films are composed of crystallites of varying size, thickness, and orientation. The characteristic Raman peaks (especially FWHM of E$_g$ peak) correlate well with the measured electronic sheet resistance across many orders of magnitude and are reasonably well in line with the extracted effective charge carrier mobility. Our PtSe$_2$ samples exhibit both positive and negative piezoresistive gauge factors, and show signs of both metallic and semiconducting behavior. The sign and magnitude of the GF are affected by both the material structure and the total film thickness. Our results suggest that the properties of PtSe$_2$ films can be tailored by controlling their nanocrystalline structure, which requires great control of the TAC process. For example, applications in nanoelectromechanical systems such as strain, pressure or acceleration sensors or microphones would benefit from high negative piezoresistive gauge factors, such as demonstrated here in very thin PtSe$_2$ films. We conclude that scalable and manufacturable TAC growth processes can be tuned to design application-specific PtSe$_2$ and other 2D material films.

## 4. Experimental Section

*Material synthesis:* Platinum (Pt) layers of varying thickness were deposited (sputtered or evaporated) onto the centimeter-sized $SiO_2$ or quartz growth substrates. TAC was used to convert the initial metal layers forming $PtSe_2$ as published earlier.[1,8]

The film thicknesses were either determined from the initially deposited layer of Pt prior to selenization (by multiplication with the expansion factor[23]), by means of AFM, or from the TEM analysis.

*Raman characterization:* The Raman measurements were performed using a WITec alpha 300R system with a 532 nm laser and a 1800 g/mm grating. The laser power was set to 300 µW. The lateral resolution of the system was limited to 300 nm. The FWHM was extracted from Lorentzian fits of the Raman peaks. For a statistical analysis, area scans were performed and the position and FWHM were extracted from Gaussian histogram fits. Calibration was done using the Si peak at 520 cm$^{-1}$.

*XPS characterization:* For XPS measurements, a PHI VersaProbe III instrument equipped with a micro-focused monochromated Al Kα source (1486.6 eV) and dual beam charge neutralization was used. High-resolution scans of the individual core-orbitals of interest including platinum (Pt), selenium (Se), carbon (C), and oxygen (O) were obtained. The binding energy was referenced to the Pt 4f 7/2 level and set to 73.65 eV. The common practice of using the C 1s level for reference was avoided due to the strong Kapton signal overlapping the adventitious C peak. Choosing the Pt 4f 7/2 level energy of 73.65 eV as fixed value is in agreement with comparable $PtSe_2$ films on other substrates were the adventitious C reference could be used.

*TEM sample preparation and characterization:* TEM lamellae were prepared using a focused ion beam (FIB) technique employing two different FEI Dual Beam Helios NanoLab systems causing different quality of final lamellae. The lamellae of samples 1 and 2A were thinner, which is why the final images look clearer. The annular bright field and annular dark field scanning transmission electron microscopy (ABF and ADF STEM) was carried out on an FEI Titan 80-300 probe Cs-corrected TEM operated at 200 kV.[64]

*Device fabrication and characterization:* The PtSe$_2$ films of material batches 1, 2, and 3 were transferred from their growth substrates onto Si/SiO$_2$ (90 nm) substrates using a potassium hydroxide (KOH) based wet transfer technique. A supporting layer of poly(methyl methacrylate) (PMMA) was applied to the PtSe$_2$ films and the film was scratched to provide access to the underlying oxide layer to a few drops of a KOH solution. After delamination, the PtSe$_2$/PMMA films remained floating on the surface of DI water from where they were transferred onto the final substrates by a fishing technique. The samples were dried in air and the PMMA support layer was then removed in acetone and IPA. For material batch 4, device fabrication was done on the growth substrates, which already were the same Si/SiO$_2$ (90 nm) substrates. Transfer length method (TLM) and six-port Hall bar structures were then fabricated using contact lithography and a CF$_4$/O$_2$-based reactive ion etching (RIE) process. In the first lithography step, the self-aligned edge contacts[51] were fabricated by first etching the PtSe$_2$ under the contact pads and then sputtering Ni and Al, followed by a lift-off process. Afterwards, the PtSe$_2$ channels were defined in a second lithography and etching step. The channel width and length of the Hall bar structures ($W$ and $L_{inner}$) were measured after device fabrication for each sample individually by optical microscopy.

*Hall measurements:* A bias current $I_C$ was applied to the outer contacts of the six-port Hall bar devices and the Hall voltage $V_H$ between two opposite inner contacts was measured over time while the magnetic field $B$ was switched on and off (SI **Figure S9**a). From the Hall effect measurements, the current and voltage related sensitivities $S_I$ and $S_V$ can be extracted according to **Equation (7)** and **(8)**,

$$S_I = \frac{1}{I_C} \left| \frac{\partial V_H}{\partial B} \right| \tag{7}$$

$$S_V = \frac{1}{V_C} \left| \frac{\partial V_H}{\partial B} \right| \tag{8}$$

where $V_C$ is the voltage between the two outer contacts resulting from the current bias. The sheet charge carrier density $n_\square$ and the effective Hall mobility $\mu_H$ can then be extracted according to **Equations (9)** and **(10)**,[65,66]

$$n_\square = \frac{1}{S_I\, e} \tag{9}$$

$$\mu_H = S_V \frac{L}{W} \tag{10}$$

where $e$ is the elementary charge, $L \approx 40$ µm is the total channel length, and $W \approx 10$ µm is the channel width. A lock-in amplifier (Stanford Research Systems Model SR830 DSP Lock-In Amplifier) was used to provide a sine wave signal with an amplitude of 5 V and a frequency of approximately 45 Hz as bias on the outer contacts. For the integration of the measured Hall signal, a time constant of 300 ms was chosen. The magnetic field was switched between -57.6 mT and +57.6 mT to maximize the output signal.

*Strain gauge fabrication and characterization:* Large metal contacts (Ni) were sputtered onto pieces of flexible polyimide foil (Kapton) using a shadow mask. The PtSe$_2$ films were then transferred from their growth substrates onto the polyimide foil with the same technique as described above, resulting in unpatterned strain gauges of millimeter size with bottom contacts. The PMMA from the transfer was removed from the PtSe$_2$ but a new layer of PMMA was spin-coated and baked at 115 °C for 10 minutes after the transfer to protect the devices from mechanical stress during the following steps. Electrical measurements showed that the additional PMMA layer did not significantly affect the resistance of the strain gauges. All samples were then glued to the steel beam (300 mm × 30 mm × 3 mm) with a distance of 200 mm to the loading point. The applied mechanical strain was calculated and simulated depending on the attached load in the same way as done previously.[2] The measurement set-up is shown in SI **Figure S12**, including a video recording of a measurement.

**Supporting Information**

Supporting Information is available from the Wiley Online Library or from the author.

**Acknowledgements**

This work has received funding from the German Ministry of Education and Research BMBF under grant No 16ES1121 (NobleNEMS) and from the European Union's Horizon 2020 research and innovation programme under grant agreements 829035 (QUEFORMAL), 825272 (ULISSES), Graphene Flagship Core 3 (881603).